\documentclass[10pt,aps,prx,twocolumn,notitlepage,showpacs,superscriptaddress, 
]{revtex4-1}
\usepackage{graphicx}
\usepackage{amssymb,amsmath,bm}
\usepackage{footnote}
\usepackage[colorlinks, linkcolor=blue,anchorcolor=blue,citecolor=blue,urlcolor=blue]{hyperref}
\newcommand{\be}{\begin{equation}}
\newcommand{\ee}{\end{equation}}

\renewcommand{\b}[1]{{\boldsymbol{#1}}}
\newcommand{\tr}{\mathop{\mathrm{Tr}}}
\newcommand{\bsigma}{\boldsymbol{\sigma}}
\newcommand{\re}{\mathop{\mathrm{Re}}}
\newcommand{\im}{\mathop{\mathrm{Im}}}

\begin{document}
\title{Analytical approach for the Mott transition in the Kane-Mele-Hubbard model}
\author{Joel Hutchinson}
\affiliation{CPHT, CNRS, Ecole Polytechnique, Institut Polytechnique de Paris, Route de Saclay, 91128 Palaiseau, France} 

\author{Philipp W. Klein}
\affiliation{CPHT, CNRS, Ecole Polytechnique, Institut Polytechnique de Paris, Route de Saclay, 91128 Palaiseau, France} 

\author{Karyn Le Hur}
\affiliation{CPHT, CNRS, Ecole Polytechnique, Institut Polytechnique de Paris, Route de Saclay, 91128 Palaiseau, France} 
\date{\today}
\begin{abstract}
The description of interactions in strongly-correlated topological phases of matter remains a challenge. 
Here, we develop a stochastic functional approach for interacting topological insulators including both charge and spin channels. We find that the Mott transition of the Kane-Mele-Hubbard model may be described by the variational principle with one equation.  We present different views of this equation from the electron Green's function, the free-energy and the Hellmann-Feynman theorem. In particular, we show the stability of the transition line towards fluctuations, in good agreement with numerical results. The band gap remains finite at the transition and the Mott phase is characterized by antiferromagnetism in the $x-y$ plane. The interacting topological phase is described through a $\mathbb{Z}_2$ number related to helical edge modes. Our results then show that improving stochastic approaches can give further insight on the understanding of interacting phases of matter.
\end{abstract}
\maketitle

\section{Introduction}

The quantum spin Hall insulator (QSHI) is a topological system of spinful fermions that preserves time reversal symmetry and similarly as a spin liquid develops short-range magnetism. It is gapped in the bulk, but has a Kramers pair of helical edge modes, one corresponding to each spin. As long as time-reversal symmetry is maintained, the modes will cross at time-reversal invariant momenta in the energy band structure. These modes are protected from backscattering off of non-magnetic impurities by a $\mathbb{Z}_2$ invariant, making such systems promising candidates for quantum electronics~\cite{bernevig2006}. The effect has been observed in ${\rm HgTe}$ quantum wells \cite{Wurzburg}, in three-dimensional Bismuth materials~\cite{Hasan}, and proposals exist for demonstrating it in ultracold atoms~\cite{kennedy2013}.

The canonical model for the QSHI is the Kane-Mele model~\cite{KM}, in which the topological phase is induced by spin-orbit coupling through the next-nearest neighbour hopping term on a hexagonal lattice. While the edge modes are protected from single-particle scattering by the $\mathbb{Z}_2$ invariant, their robustness to two-particle scattering requires more careful analysis. It has been shown that this phase is stable to weak interactions~\cite{xu2006}. More generally, at half filling, the effect of interactions has been seen in several studies, typically through the addition of an on-site Hubbard interaction to the Kane-Mele model~\cite{young2008,KMH1,Rachel_2018,KMH2,Hohenadler_2011,Hohenadler_2012,Zheng_2011,Laubach_2014,Zeng_2017}.  These studies show the existence of a magnetic phase at strong coupling which destroys the topological order. 
On the other hand, a simple analytical description of the Mott transition line in agreement with numerical methods \cite{Hohenadler_2012, KMH2} remains to be developed linked to the development of strongly-correlated materials \cite{Rachel_2018} and the tunability
of interactions in ultra-cold atoms \cite{Harvard}. 

In this article, we revisit the Mott transition in the Kane-Mele-Hubbard model developing an analytic path integral stochastic approach keeping both the charge and spin channels. 
We show that with a decoupling scheme that preserves the symmetry of the Hubbard interaction and with a variational approach, the transition line --- which is described through one equation (\ref{eq:Uc}) --- shows quantitative agreement with numerical methods such as quantum Monte Carlo, and dynamical mean field theory~\cite{Hohenadler_2012, KMH1}. We also verify the protection of the transition line towards gaussian fluctuations. \\

 We study the half-filled Kane-Mele model~\cite{KM}, 
with a repulsive on-site Hubbard interaction such that in real space the Hamiltonian reads~\cite{KM,KM2}:
\begin{eqnarray}
\mathcal{H}&=& -t_1\sum_{\left\langle i,j\right\rangle }\sum_{\alpha} c_{i\alpha}^{\dagger} c_{j\alpha} -i t_2\sum_{\left\langle\left\langle i,j \right\rangle\right\rangle}\sum_{\alpha, \beta}  \nu_{ij} c_{i\alpha}^{\dagger}\sigma^z_{\alpha\beta} c_{j\beta}\nonumber\\
&&+U\sum_i n_{i\uparrow}n_{i\downarrow}.
\label{eq_km}
\end{eqnarray}
Here, $c_{i\alpha}^{\dagger}$ and $c_{i\alpha}$ denote fermionic creation and annihilation operators, respectively.
The sum over $\left\langle i,j\right\rangle$ refers to nearest-neighbors with hopping amplitude $t_1$, while the sum over $\left\langle\left\langle i,j\right\rangle\right\rangle$ refers to next-nearest neighbors with hopping amplitude $t_2$ and spin-orbit coupling $\nu_{ij}=\pm 1$ depending on whether going from $i$ to $j$ means moving clockwise or counter-clockwise around the plaquette. 
Lastly, $\sigma^z$ denotes the third Pauli matrix in spin space with components $\alpha,\beta\in\left\lbrace\uparrow,\downarrow\right\rbrace$. 
The non-interacting model at $U=0$ reveals a bulk band insulator with two degenerate counter-propagating helical edge modes associated with the spin components $\uparrow$ and $\downarrow$. The gap grows linearly with $t_2$ until $t_2\approx0.2t_1$, at which point it remains constant. It has band Chern numbers of $+1$ and $-1$ that can be associated to the spin components $\uparrow$ and $\downarrow$, respectively. While the total Chern number is zero, Kane and Mele showed that the model shows a $\mathbb{Z}_2$ topological index \cite{KM,KM2}.

When electron-electron interactions are added to the model, the topological band insulator is challenged by correlation physics, and is no longer exactly solvable. The addition of an on-site Hubbard interaction to the Kane-Mele model has been studied within various approximations in the past decade~\cite{Rachel_2018,KMH1,KMH2,Griset_2012,Hamad_2016,Lee_2011,Mardani_2011,Soriano_2010,Hohenadler_2011,Hohenadler_2012,Hung_2013,Laubach_2014,MENG_2013,Yamaji_2011,Yu_2011,Zheng_2011,Zheng_2011}. The model comprises two phases. First, up to some critical interaction strength $U_c \gtrsim t_1$ the topological band insulator is stable towards electron-electron interactions~\cite{Rachel_2018}. Upon reaching the critical $U_c$, the system transitions to a magnetically ordered phase (spin density wave). In this phase, the system prefers to magnetically order in the $x-y$ plane only. To the best of our knowledge, the precise location of the transition line defined through $U_c$ and properties of the Mott transition remain to be addressed, supporting numerical findings  \cite{KMH2,Hohenadler_2011,Hohenadler_2012,Zheng_2011,Laubach_2014,Zeng_2017}. Understanding properties of the Mott transition in the bosonic Kane-Mele-Hubbard model also requires the introduction of numerical methods \cite{Harvard}. Therefore, we intend now to develop a stochastic approach analytically. We have recently shown that such a methodological approach gives quantitative results for the Mott transition in the Haldane model \cite{StochasticHaldane}. For the Kane-Mele model, we generalize this stochastic scheme including charge and spin channels and we derive analytically the transition line to the Mott state. 

The article is organized as follows. In Sec. \ref{sec:method}, we introduce the methodological steps related to the stochastic approach. In Sec. \ref{Results}, we derive the phase diagram from the variational principle. In Sec. \ref{Fluctuations}, we show the robustness of the analytical results towards
fluctuations. In Sec. \ref{sec:HF}, we address the energetics of the interacting model. In Sec. \ref{conclusion}, we summarize the main findings.

\section{Stochastic Method}\label{sec:method}

We first write the interaction in Eq. (\ref{eq_km}) as the square of a bilinear operator. For spin-1/2 fermions, the space of such operators is spanned by the Pauli matrices along with identity matrix, so one can generically write  
\begin{align}
		U \sum_{i} n_{i\uparrow} n_{i\downarrow} &= U \sum_{i} c_{i\uparrow}^{\dagger} c_{i\uparrow} c_{i\downarrow}^{\dagger} c_{i\downarrow} \\ 
										&= U \sum_{i,r} \eta_r S_i^rS_i^r \label{equ_decomposition_quartic_term}
\end{align}
where $S^r_i\equiv c^\dagger_{i\alpha}\sigma^r_{\alpha\beta}c_{i\beta}$, $i$ denotes a lattice site and $r\in\left\lbrace 0,x,y,z\right\rbrace$ with the general properties that $\eta_x=\eta_y$ and $\eta_z=-\eta_0$. Previous mean-field studies kept only the $S^0_i$ and $S^z_i$ channels with $\eta_0=1/4=-\eta_z$ \cite{KMH1}. In this article, we include the $x$ and $y$ channels as well. In the $t_2\rightarrow0$ limit, the model retains the full spin rotation symmetry, and this should be reflected in our choice of interaction representation \cite{schulz_1994}. 
The symmetric decomposition $\eta_0=1/8$, $\eta_{x,y,z}=-1/8$ yields the largest symmetry group $SU(2)$ of the interaction such that

\be
\mathcal{H}_U=\frac{U}{8}\sum_i\b{S}_i\cdot \b{S}_i +\frac{U}{4}\sum_i(n_{i\uparrow}+n_{i\downarrow}),\label{eq:int}
\ee
using the Minkowski inner product~\footnote{We use bold-font ($\b{S}$ and $\b{\phi}$)) to refer to four-vectors with a Minkowski inner product, and hats ($\vec{S}$ and $\vec{\phi}$) to refer to three-vectors with a Euclidean inner product.}, $\b{S}_i\cdot \b{S}_i=(S^0_i)^2-(S^x_i)^2-(S^y_i)^2-(S^z_i)^2$. Our first goal is to apply the Hubbard-Stratonovich transformation with all four channels weighted equally to obtain the phase diagram for this model. This symmetric choice of $\eta$'s is unbiased towards any particular type of order. 

We start with the action
\be
\mathcal{S}[\psi^\dagger,\psi]=\mathcal{S}_0[\psi^\dagger,\psi]+\int^\beta_0d\tau\frac{U}{8}\sum_i\b{S}_i\cdot \b{S}_i,
\ee
where $\psi=(c_{A\uparrow},c_{B\uparrow},c_{A\downarrow},c_{B\downarrow})^T$ and $S_0[\psi^\dagger,\psi]$ is the action for the non-interacting Kane-Mele model (including the chemical potential shift $\frac{U}{4}$ from Eq.~\eqref{eq:int}).

\subsection{Hubbard-Stratonovitch Action}

We insert a resolution of the identity for four independent bosonic dimensionless fields $\phi^r_i$ corresponding to each $S_i^r$ channel and renormalize by the constant determinant. The resulting path integral is
\begin{eqnarray}
\mathcal{Z}&=&\int\Pi_r\mathcal{D}\phi^r\int\mathcal{D}\psi^\dagger\mathcal{D}\psi\exp\bigg(-\mathcal{S}[\psi^\dagger,\psi]\nonumber\\
&&-2U\int_0^\beta d\tau\sum_{i,r}\phi_i^r\phi_i^r\bigg),
\end{eqnarray}
where $\Pi_r$ refers to the product on $r=0,x,y,z$.
The following linear transform in the bosonic fields, $\phi_i^0 \rightarrow \frac{i}{2}\phi_i^0+\frac{i}{4}S^0_i$ and $\phi_i^{x,y,z}\rightarrow\frac{1}{2}\phi_i^{x,y,z}+\frac{1}{4}S_i^{x,y,z}$,
will compensate the interaction term in $-\mathcal{S}[\psi^\dagger,\psi]$ such that
\begin{eqnarray}
\mathcal{Z}&=&\frac{i}{16}\int\Pi_r\mathcal{D}\phi^r\int\mathcal{D}\psi^\dagger\mathcal{D}\psi\exp\bigg(-\mathcal{S}_0[\psi^\dagger,\psi] \nonumber \\
&& +\frac{U}{2}\int^\beta_0d\tau\sum_{i}(\b{\phi}_i\cdot\b{\phi}_i+\b{\phi}_i\cdot\b{S}_i)\bigg),
\end{eqnarray}
where we have defined the four-vector $\b{\phi}_i\equiv(\phi_i^0,\phi_i^x,\phi_i^y,\phi_i^z)$ and used the Minkowski inner product again. The classical field equations obtained from $\frac{\delta \mathcal{S}}{\delta \phi^r_i}=0$ allows us to relate $\b{\phi}_i$ to the fermion fields 
\begin{eqnarray}
\b{\phi}_i = -\frac{1}{2}\langle \b{S}_i\rangle=-\frac{1}{2}\langle c^\dagger_{i\alpha}\b{\sigma}_{\alpha\beta}c_{i\beta}\rangle.\label{eq:phic}
\end{eqnarray}
To obtain an insulator, we fix the particle density at half-filling such that $\phi^0=-1/2$. The stochastic fields are static variables allowing us to evaluate the electron Green's function and energetics for a given fields configuration and then to apply the variational principle
to find the most favorable distribution of those variables. 

\subsection{Mean-Field Hamiltonian}

We will consider fields with the translation symmetry of the lattice. Upon Fourier transforming, this means that $\b\phi_{\b{k}s}=\sqrt{N}\delta_{\b{k},0}\b\phi_{s}$, where $N$ is the number of unit cells, $\b\phi_{s}$ are intensive constants and we have made the sublattice index $s\in\{A,B\}$ explicit. This means the mean-field Hamiltonian is diagonal in momentum space:
\onecolumngrid
\begin{equation}
\mathcal{H}_{\rm MF}(\b{k})=\begin{pmatrix}
\gamma(\b{k})+\frac{U}{2}\phi^z_{A} & -g(\b{k}) & \frac{U}{2}(\phi^x_{A}+i\phi^y_{A})& 0\\
-g^*(\b{k}) & -\gamma(\b{k})+\frac{U}{2}\phi^z_{B} & 0 & \frac{U}{2}(\phi^x_{B}+i\phi^y_{B})\\
\frac{U}{2}(\phi^x_{A}-i\phi^y_{A}) & 0 & -\gamma(\b{k})-\frac{U}{2}\phi^z_{A} & -g(\b{k})\\
0 & \frac{U}{2}(\phi^x_{B}-i\phi^y_{B}) & -g^*(\b{k}) & \gamma(\b{k})-\frac{U}{2}\phi^z_{B}
\end{pmatrix}.
\end{equation}
\twocolumngrid
The functions $\gamma$ and $g$ are defined as 
\begin{align}
 \gamma(\bm{k}) &= -2t_2\sum_p\sin(\bm{k}\cdot \bm{b}_p)\label{equ_gamma_func}\\
 g(\bm{k}) &= t_1 \sum_p\left( \cos(\bm{k}\cdot \bm{a}_p)-i \sin(\bm{k}\cdot \bm{a}_p)\right). \label{equ_g_func}
\end{align}
The nearest and next nearest neighbor displacements on the honeycomb lattice are denoted as $\bm{a}_p$ and $\bm{b}_p$ following definitions of Ref.~\cite{StochasticHaldane}.

The action is then
\begin{eqnarray}
\mathcal{S}&=&\int^\beta_0d\tau\bigg[\sum_{\b{k}}\psi^\dagger_\b{k}\bigg(\partial_\tau+\frac{U}{2}+\mathcal{H}_{\rm MF}(\b{k})\bigg)\psi_\b{k}\nonumber\\
&&-\frac{U}{2}\sum_{\b{k},s}(\b{\phi}_{\b{k}s}\cdot\b{\phi}_{-\b{k}s})\bigg].
\end{eqnarray}


We then transform the action into frequency space so that $\b\phi_{n\b{k}s}=\beta\delta_{i\omega_n,0}\b\phi_{\b{k}s}$ or equivalently $\phi^r_{n\b{k}s}=\beta\delta_{i\omega_n,0}\phi^r_{\b{k}s}$, where $\omega_n$ are fermionic Matsubara frequencies and $\b\phi_{\b{k}s}$ is independent of frequency and time:
\be
\psi(\tau)=\frac{1}{\beta}\sum_{i\omega_n}\psi_ne^{-i\omega_n\tau}.
\ee

Since the action consists of fermion bilinears, we may integrate out the fermions to get a determinant:
\begin{eqnarray}
\mathcal{Z}&=&\frac{i}{16}\int\Pi_{r,\b{k},s}d\phi^r_{\b{k}s}\det(-\beta\mathcal{G}^{-1}(i\omega_n))\nonumber\\
&&\times\exp\bigg(\frac{\beta U}{2}\sum_{\b{k},s}(\b{\phi}_{\b{k}s}\cdot\b{\phi}_{-\b{k}s})\bigg),
\end{eqnarray}
where the inverse fermion Green's function is
\be
\mathcal{G}^{-1}_{\b{q}\b{k}}(i\omega_n)=\bigg(i\omega_n-\frac{U}{2}-\mathcal{H}_{\rm MF}(\b{k})\bigg)\delta_{\b{q},\b{k}}.
\ee
Thus we have the effective Hubbard-Stratonovich action
\begin{eqnarray}
\mathcal{S}_{\rm HS}
&=& -\beta U\sum_{\b{k}}(\b{\phi}_{\b{k}}\cdot\b{\phi}_{-\b{k}})-\tr\ln(-\beta\mathcal{G}^{-1}(i\omega_n)),\label{eq:SHS}
\end{eqnarray}
where $\tr$ denotes the trace over Matsubara frequencies, momentum space, spin space and pseudospin space. We put some assumptions on the type of ordering that may occur. Superexchange induces antiferromagnetic order \cite{Rachel_2018} such that, $\phi^{x,y,z}_{\b{k}A}=- \phi^{x,y,z}_{\b{k}B}\equiv\phi^{x,y,z}_{\b{k}}$. 

\section{Results}
\label{Results}

To derive the phase diagram analytically, we vary the Hubbard-Stratonovich action with respect to the HS fields. 

\subsection{Green's Functions, Observables and Transition}

We get the saddle-point conditions
\begin{eqnarray}
\frac{\delta \mathcal{S}_{\rm HS}}{\delta \phi^r_{\b{p}}}&=&\pm2\beta U\phi^r_{-\b{p}}-\tr\bigg[\mathcal{G}(i\omega_n)\frac{\delta \mathcal{G}^{-1}(i\omega_n)}{\delta \phi_\b{p}^r}\bigg],\label{eq:Sderiv}
\end{eqnarray}
$+$ holds for $r=x,y,z$ and $-$ holds for $r=0$. Specifically,
\begin{eqnarray}
\frac{\delta \mathcal{S}_{\rm HS}}{\delta \phi^0_0}&=&-2\beta\sqrt{N} U\phi^0-\frac{U}{2\sqrt{N}}\sum_{i\omega_n,\b{k}}{\rm tr}\mathcal{G}(i\omega_n,\b{k}),\label{eq:tr1}\\
\frac{\delta \mathcal{S}_{\rm HS}}{\delta \phi^r_0}&=&2\beta\sqrt{N} U\phi^r+\frac{U}{2\sqrt{N}}\sum_{i\omega_n,\b{k}}{\rm tr}\bigg(\mathcal{G}(i\omega_n,\b{k})(\sigma^r\otimes\tau^z)\bigg),\label{eq:tr2}\nonumber\\
\end{eqnarray}
where now ${\rm tr}$ refers to the matrix trace, $r=x,y,z$, $\sigma^r$ are Pauli matrices in spin space and $\tau^z$ is the third Pauli matrix in the sublattice space. 
The Matsubara Green's function can be evaluated analytically:
\be
\mathcal{G}(i\omega_n,\b{k})=\frac{(i\omega_n-\frac{U}{2})\mathbb{I}+\mathcal{H}_{\rm MF}(\b{k})}{(i\omega_n-E_{\b{k}+})(i\omega_n-E_{\b{k}-})},\label{eq:green}
\ee

The poles of the Green's function are the quasi-particle energies
\be
E_{\b{k}\pm}=\frac{U}{2}\pm\sqrt{\epsilon_\b{k}^2+2\gamma_\b{k}\bigg(\frac{U}{2}\bigg)\phi^z+\bigg(\frac{U}{2}\bigg)^2\vec\phi\cdot\vec\phi},\label{eq:qpenergy}
\ee
where we have defined the non-interacting dispersion $\epsilon_\b{k}\equiv\sqrt{\gamma(\b{k})^2+|g(\b{k})|^2}$.
We see that the chemical potential is effectively shifted by $\frac{U}{2}$.

\begin{figure}[ht]
	\centering
	\includegraphics[width=0.88\columnwidth]{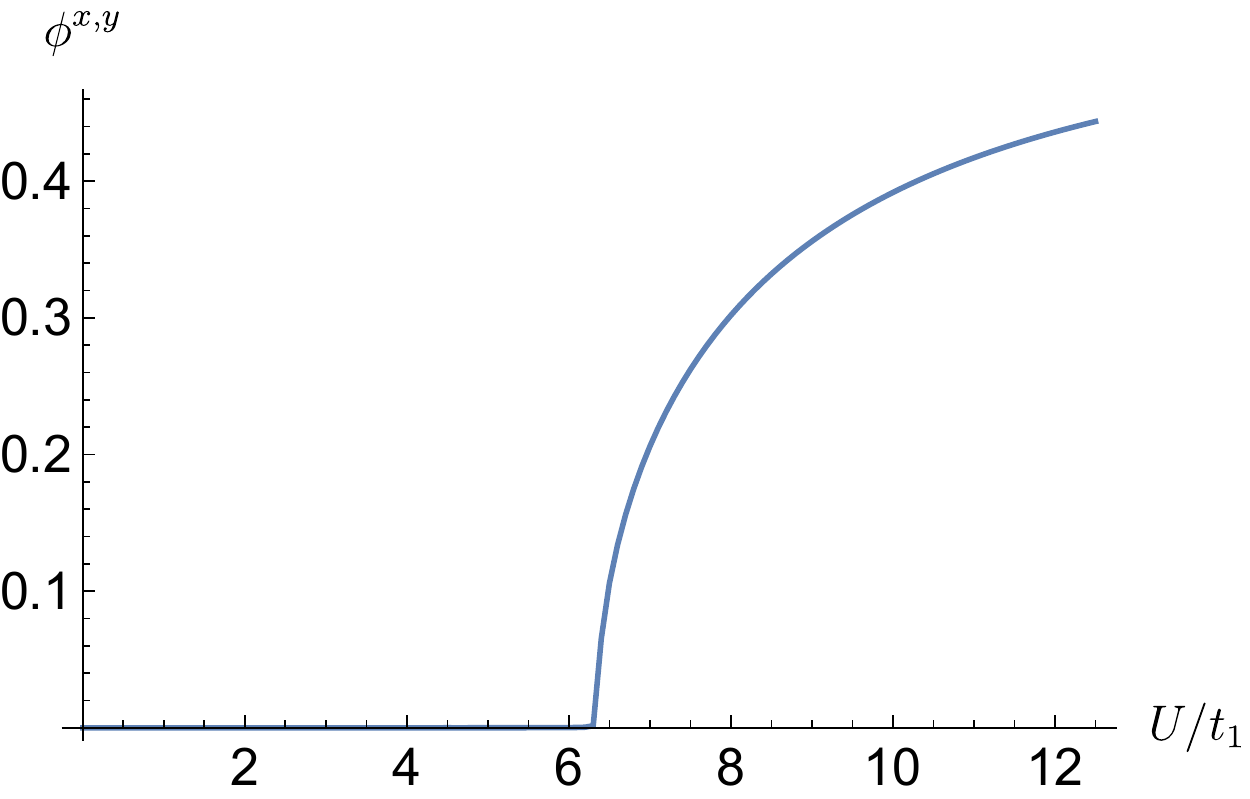}
\caption{Magnetization profile for $t_2=0.3t_1$.}\label{fig:mag}
\end{figure}

\begin{figure}[h]
	\centering
	\includegraphics[width=0.98\columnwidth]{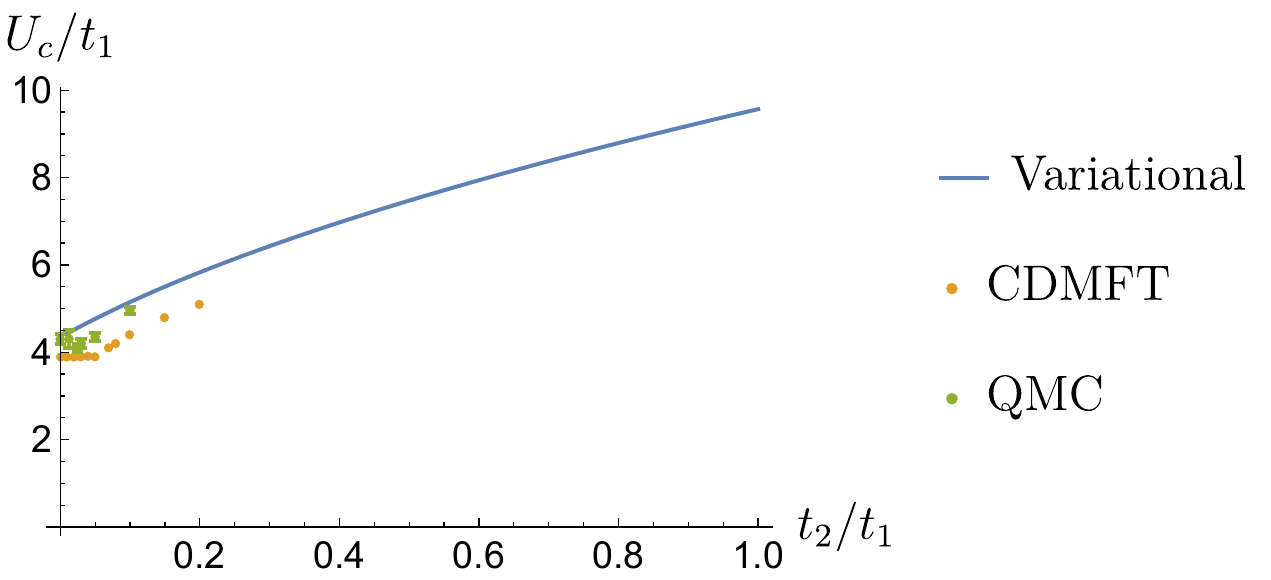}
\caption{Onset of antiferromagnetic XY order at the Mott transition versus $t_2/t_1$ from the variational stochastic approach defined through $U_c$ in Eq. (\ref{eq:phixy}) (solid blue line). This is compared to previous data from CDMFT in orange (Ref.~\cite{KMH2}) and QMC in green (Ref.~\cite{Hohenadler_2012}).}\label{fig:trans}
\end{figure}

The traces in Eq.~\eqref{eq:tr1},~\eqref{eq:tr2} are readily evaluated. At zero temperature, the saddle-point conditions $\frac{\delta\mathcal{S}_{\rm HS}}{\delta\phi^r}=0$ then yield:
\be
\phi^{x,y}=\frac{U\phi^{x,y}}{4N}\sum_\b{k}\frac{1}{\sqrt{\epsilon_\b{k}^2+2\gamma_\b{k}(\frac{U}{2})\phi^z+(\frac{U}{2})^2\vec\phi\cdot\vec\phi}}.\label{eq:phixy}
\ee
There is a second-order transition in this magnetic order parameter as shown in Fig.~\ref{fig:mag}.
Linearizing $\vec{\phi}$ around the zero vector, one finds the critical coupling:
\be
\frac{1}{U_c^{x,y}}=\frac{1}{4N}\sum_\b{k}\frac{1}{\epsilon_\b{k}},\label{eq:Uc}
\ee
shown in Fig.~\ref{fig:trans}. This result shows remarkable quantitative agreement with quantum Monte Carlo (QMC) and cluster dynamical mean field theory (CDMFT) for small $t_2$~\cite{Hohenadler_2012, KMH2, sorella2012}.

Meanwhile, the saddle-point condition for $\phi^z$ is
\be
\phi^{z}=\frac{1}{2N}\sum_\b{k}\frac{\gamma_\b{k}+\frac{U}{2}\phi^z}{\sqrt{\epsilon_\b{k}^2+2\gamma_\b{k}(\frac{U}{2})\phi^z+(\frac{U}{2})^2\vec\phi\cdot\vec\phi}}.\label{eq:phiz}
\ee
Linearizing $\vec{\phi}$ about the zero vector, and noting that $\gamma_\b{k}$ is odd under inversion also gives the critical coupling
\be
\frac{1}{U_c^z}=\frac{1}{4N}\sum_\b{k}\frac{|g_\b{k}|^2}{\epsilon_\b{k}^3}.
\ee
Since $|g_\b{k}|^2<\epsilon_\b{k}^2$ for all $t_2>0$, we see that $U_c^{x,y}<U_c^z$ except at $t_2=0$ at which point the transition lines are identical and the full $SU(2)$ symmetry is restored. Thus as we approach from the normal state, the spins will first order antiferromagnetically in the $x-y$ plane. In fact, we can go a step further. For $t_2>0$, it turns out that 
$\phi^z$ must vanish for any $U$. We can see this by combining Eqs.~\eqref{eq:phixy} and~\eqref{eq:phiz} to get
\be
\phi^z=\phi^z+\frac{1}{2N}\sum_\b{k}\frac{\gamma_k}{\sqrt{\epsilon_\b{k}^2+2\gamma_\b{k}(\frac{U}{2})\phi^z+(\frac{U}{2})^2\vec\phi\cdot\vec\phi}}. 
\ee
In order for the sum to vanish at finite $t_2$, the denominator must be invariant under inversion. This only happens for $\phi^z=0$, so we confirm that the ordering strictly takes place in the plane for all $U$, in agreement with quantum Monte Carlo and strong-coupling results~\cite{Hohenadler_2012, KMH1}. As long as $\phi^z=0$, then we verify from Eq. (\ref{eq:qpenergy}) that the gap does not close at the phase transition, but is uniformly renormalized by $(U/2)^2\vec{\phi}\cdot\vec{\phi}$.

\subsection{Transition from the Hellmann-Feynman theorem} 

Here we note that the transition line may also be computed directly from the quasi-particle energies Eq.~\eqref{eq:qpenergy}, using the Hellmann-Feynman theorem~\cite{guttinger1932, hellmann1937}:
\be
\frac{dE_{\rm gs}}{d\lambda}=\frac{d}{d\lambda}\langle \psi_{\rm gs}|\mathcal{H}|\psi_{\rm gs}\rangle,\label{eq:HellFeyn}
\ee
where $E_{\rm gs}$ is the energy of the ground state $|\psi_{\rm gs}\rangle$ and $\lambda$ is any parameter of the system. At half-filling, if we choose $\lambda=\phi^x$ for example, the left-hand side is
\begin{eqnarray}
\frac{dE_{\rm gs}}{d\phi^x}&=&
-\frac{U^2}{2}\sum_\b{k}\frac{\phi^x}{\sqrt{\epsilon_\b{k}^2+2\gamma_\b{k}(\frac{U}{2})\phi^z+(\frac{U}{2})^2\vec\phi\cdot\vec\phi}}.\label{eq:lhs}
\end{eqnarray}
The right-hand side is 
\begin{eqnarray}
\frac{d}{d\phi^x}\langle \psi_{\rm gs}|\mathcal{H}|\psi_{\rm gs}\rangle&=&
\frac{U}{2}\bigg\langle\sum_\b{k}\psi^\dagger_\b{k}(\sigma^x\otimes\tau^z)\psi_\b{k}\bigg\rangle\\
&=&-2UN\phi^x.\label{eq:rhs}
\end{eqnarray}
Setting \eqref{eq:lhs}=\eqref{eq:rhs} gives Eq.~\eqref{eq:phixy}. Note that this value of the (mean-field) transition line derives precisely from the channel-symmetric decomposition in Eq~\eqref{equ_decomposition_quartic_term}. We also address a comparison with the Hartree-Fock method in Sec.~\ref{sec:HF}.

\section{Fluctuations} 
\label{Fluctuations}

Here, we justify that the transition line in Fig.~\ref{fig:trans} is stable towards Gaussian fluctuation effects from the behavior of the polarization bubble in the charge and spin channels. Indeed, we verify below that fluctuations in the charge and spin channels are suppressed in the limit of long-wavelengths and low-energy due to a gap in the spin-wave dispersions for small $t_2$. In addition, we find that the Goldstone mode associated to phase fluctuations in the $xy$ plane does not modify the transition line. This result comes from the fact that taking into account gaussian fluctuations, the dispersion of this mode keeps a similar form as for the graphene band structure~\cite{gonzalez2008}.

We start by taking a second derivative of the Hubbard-Stratonovich action with respect to the fields, which gives
\begin{eqnarray}
\frac{\delta^2 \mathcal{S}_{\rm HS}}{\delta\phi^{s'}_{\b{p}'}\delta \phi^s_{\b{p}}}
&=&\pm2\beta U\delta_{ss'}\delta_{\b{p}',-\b{p}}+\tr\bigg(\mathcal{G}\frac{\delta\mathcal{G}^{-1}}{\delta\phi^{s'}_{\b{p}'}}\mathcal{G}\frac{\delta\mathcal{G}^{-1}}{\delta\phi^{s}_{\b{p}}}\bigg),\nonumber\\
\end{eqnarray}
where $+$ holds for $s=x,y,z$, $-$ holds for $s=0$, and we used the fact that $\frac{\delta}{\delta\phi}(\mathcal{G}\mathcal{G}^{-1})=0\Rightarrow\frac{\delta\mathcal{G}}{\delta\phi}=-\mathcal{G}\frac{\delta\mathcal{G}^{-1}}{\delta\phi}\mathcal{G}$, and that $\mathcal{G}^{-1}$ depends linearly on $\phi$. Thus to second order in fluctuations $\delta\phi^s_\b{p}\equiv\phi^s_\b{p}-\langle\phi^s_\b{p}\rangle$ about the mean-field values, the Hubbard-Stratonovich action is 
\begin{eqnarray}
\mathcal{S}_{\rm{HS}}&\approx& S(\{\langle\phi\rangle_{\rm MF}\})+\beta U\sum_{ss'}\sum_{\b{p},i\omega_n}\delta\phi^{s}_{\b{p}}L^{-1}_{ss'}(\b{p},i\omega_n)\delta\phi^{s'}_{-\b{p}},\nonumber\\
\label{eq:S2}
\end{eqnarray}
where
\be
L^{-1}_{ss'}(\b{p},i\omega_n)\equiv\pm\delta_{ss'}+\frac{U}{8}\Pi^{ss'}(\b{p},i\omega_n)
\ee
is the the inverse fluctuation propagator. The polarization bubbles $\Pi^{ss'}(\b{p},i\omega_n)$ may couple to any of the four fluctuation channels defined by the mean-field decomposition and can be computed analytically for a low-energy continuum model relevant for the small $t_2$ regime. 
\subsection{Continuum model for fluctuation propagators}
We consider the continuum Hamiltonian
\be
\mathcal{H}=v_F\int d^2\b{k} \psi^\dagger_{t\b{k}}\bigg(\b{\tau}^t\cdot\b{k}+\frac{\lambda}{v_F}\tau^t_z\sigma_z+\frac{1}{v_F}\b{m}\cdot\bsigma\tau_z\bigg)\psi_{t\b{k}},
\ee
where $v_F=3t_1/2$ is the Fermi velocity, $\lambda=3\sqrt{3}t_2$,  $t$ is the valley index such that, $\b{\tau}^t=(\tau_x,\tau_y,\pm\tau_z)$ are Pauli matrices in the sublattice space with $+$ for the $K'$ valley and $-$ for the $K$ valley. Likewise, $\bsigma=(\sigma_z, \sigma_y, \sigma_z))$ are Pauli matrices in the spin space. We include a term with the mean fields $\b{m}=(m_x,m_y)=\frac{U}{2}(\phi^x,\phi^y)$ from the in-plane antiferromagnetic order, which allows us to study fluctuations from the ordered side of the transition, but neglect the chemical potential term which will not affect the result since the gap remains open. We focus on intra-valley scattering relevant for small momentum transfer. The inverse Matsubara Green's function is 
\be
\mathcal{G}^{-1}_t(\b{k},ik_n)= ik_n-v_F\b{k}\cdot\b{\tau}-\lambda\tau_z^t\sigma_z-\b{m}\cdot\bsigma\tau_z,
\ee
which may be inverted to give,
\be
\mathcal{G}_t(\b{k},ik_n)=\frac{1}{2}\sum_{\alpha=\pm}\frac{\mathbb{I}+\alpha\hat{\gamma}^t_\b{k}}{ik_n-\alpha E_\b{k}},
\ee
where $E_\b{k}=\sqrt{(v_Fk)^2+\lambda^2+m^2}$ is the quasi-particle energy and 
\be
\hat{\gamma}^t_\b{k}\equiv\frac{v_F\b{k}\cdot\b{\tau}+\lambda\tau^t_z\sigma_z+\b{m}\cdot\bsigma\tau_z}{E_\b{k}}.
\ee

The polarization functions are given by
\begin{eqnarray}
\Pi^{ss'}(\b{q},i\omega_n)&=&\frac{1}{\beta}\sum_t\int\frac{d^2\b{k}}{(2\pi)^2}\int\frac{d(ik_n)}{2\pi}\nonumber\\
&\times&\frac{1}{4}\sum_{\alpha\alpha'}\frac{f^{ss'}_t(\b{k},\b{q})}{(ik_n+i\omega_n-\alpha E_{\b{k+q}})(ik_n-\alpha'E_\b{k})}\nonumber\\
\end{eqnarray}

where 
\be
f^{ss'}_t(\b{k},\b{q})\equiv{\rm tr}[(\mathbb{I}+\alpha\hat{\gamma}^t_{\b{k+q}})\mathcal{M}^s(\mathbb{I}+\alpha'\hat{\gamma}^t_{\b{k}})\mathcal{M}^{s'}],
\ee
with $s,s'\in\{0,x,y,z\}$, $\mathcal{M}^0=\mathbb{I}$ and $\mathcal{M}^{x,y,z}=\tau_z\sigma_{x,y,z}$. Evaluating the Matsubara sum at half-filling and zero temperature fixes $\alpha'=-\alpha$, and
\begin{eqnarray}
\Pi^{ss'}(\b{q},i\omega_n)
&=&\sum_t\int\frac{d^2\b{k}}{(2\pi)^2}\frac{1}{4}\sum_{\alpha}\frac{f^{ss'}_t(\b{k},\b{q})\alpha}{i\omega_n-\alpha (E_{\b{k+q}}+E_\b{k})}.\nonumber\\
\label{eq:piint}
\end{eqnarray}
The imaginary part of the retarded polarization is then
\begin{eqnarray}
\im\Pi^{ss'}(\b{q},\omega)&=&-\sum_t\int\frac{d^2\b{k}}{(2\pi)^2}\frac{1}{4}\sum_{\alpha}f^{ss'}_t(\b{k},\b{q})\alpha\pi\nonumber\\
&\times&\delta(\omega-\alpha (E_{\b{k+q}}+E_\b{k})).\label{eq:impi}
\end{eqnarray}
Noting that $\Pi^{ss'}(\b{q},-\omega)=\Pi^{ss'}(\b{q},\omega)^*$, we may study the region $\omega>0$ without loss of generality. In this case, the only contribution is from $\alpha=+1$. We use the delta-function to take care of the angular integral. Defining $\theta$ as the angle between $\b{k}$ and $\b{q}$, we have
\be
\delta(\omega-(E_{\b{k+q}}+E_\b{k}))=\frac{E_{\b{k+q}}}{v_F^2kq\sin\theta}\delta(\theta-\theta_0),
\ee
where $\theta_0$ is the zero of the delta-function argument, i.e. the solution of the equation 
\be
2v_F^2kq\cos\theta_0=\omega(\omega-2E_\b{k})-q^2.\label{eq:cos}
\ee
Only the solutions corresponding to a real angle $\theta_0$ will have support in the integration region. We will enforce this later by restricting the integrand of \eqref{eq:impi} to be real:
\be
\im\Pi^{ss'}(\b{q},\omega)=-\sum_t\int_0^\infty\frac{dk k}{16\pi}\frac{f^{ss'}_t(\b{k},\b{q})E_{\b{k+q}}}{v_F^2kq\sin\theta_0}\bigg|_{\theta=\theta_0\in\mathbb{R}}.
\ee
We now consider all the non-zero contributions from $f^{ss'}_t(\b{k},\b{q})$. First we note that $f_t^{xz}(\b{k},\b{q})\sim f_t^{yz}(\b{k},\b{q})\sim t\lambda$, and therefore vanishes in the sum on $t$. Note that this is not true in the case $\lambda=0$ where there is an additional contribution from $\phi^z\neq0$. The remaining off-diagonal terms are
\begin{eqnarray}
f^{0i}_t(\b{k},\b{q})&=&4m_i\bigg(\frac{E_\b{k}-E_{\b{k+q}}}{E_\b{k}E_\b{k+q}}\bigg)=-f^{i0}_t(\b{k},\b{q}),\\
f^{xy}_t(\b{k},\b{q})&=&-\frac{8m_xm_y}{E_\b{k}E_\b{k+q}}=f^{yx}_t(\b{k},\b{q})
\end{eqnarray}
for $i\in\{x,y\}$. The diagonal terms are
\begin{eqnarray}
f^{ss}_t(\b{k},\b{q})&=&4\bigg[1-{\rm sgn}(s)\bigg(\frac{v_F^2\b{k}\cdot(\b{k+q})+\tilde{m}^2}{E_\b{k}E_\b{k+q}}\bigg)\bigg]\nonumber\\
&&-\frac{8\tilde{m}_s^2}{E_\b{k}E_\b{k+q}},
\end{eqnarray}
where we have defined the generalized vector $\tilde{\b{m}}\equiv(0,m_x,m_y,\lambda)$ associated to the gap in each channel, and ${\rm sgn}(s)$ is $+$ for $s=0$ and $-$ for $s=x,y,z$.

\begin{figure}[h]
		\centering
		\includegraphics[width=0.98\columnwidth]{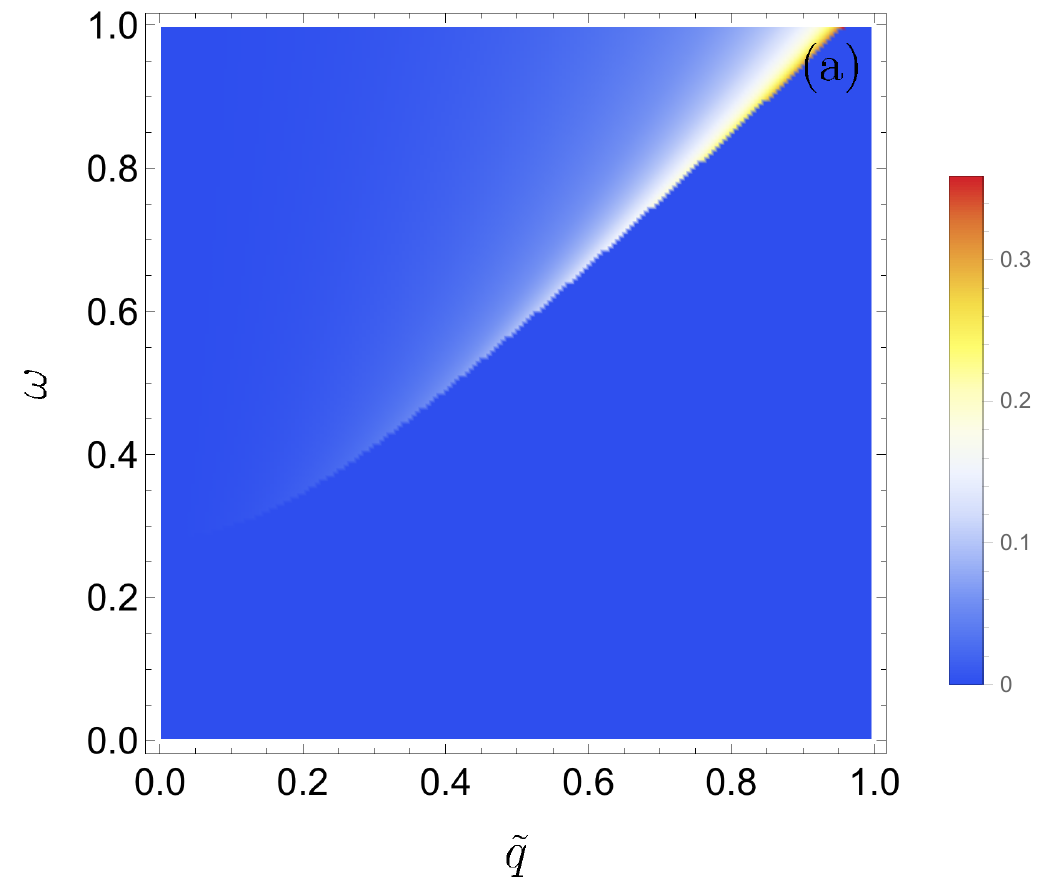}
		\includegraphics[width=0.98\columnwidth]{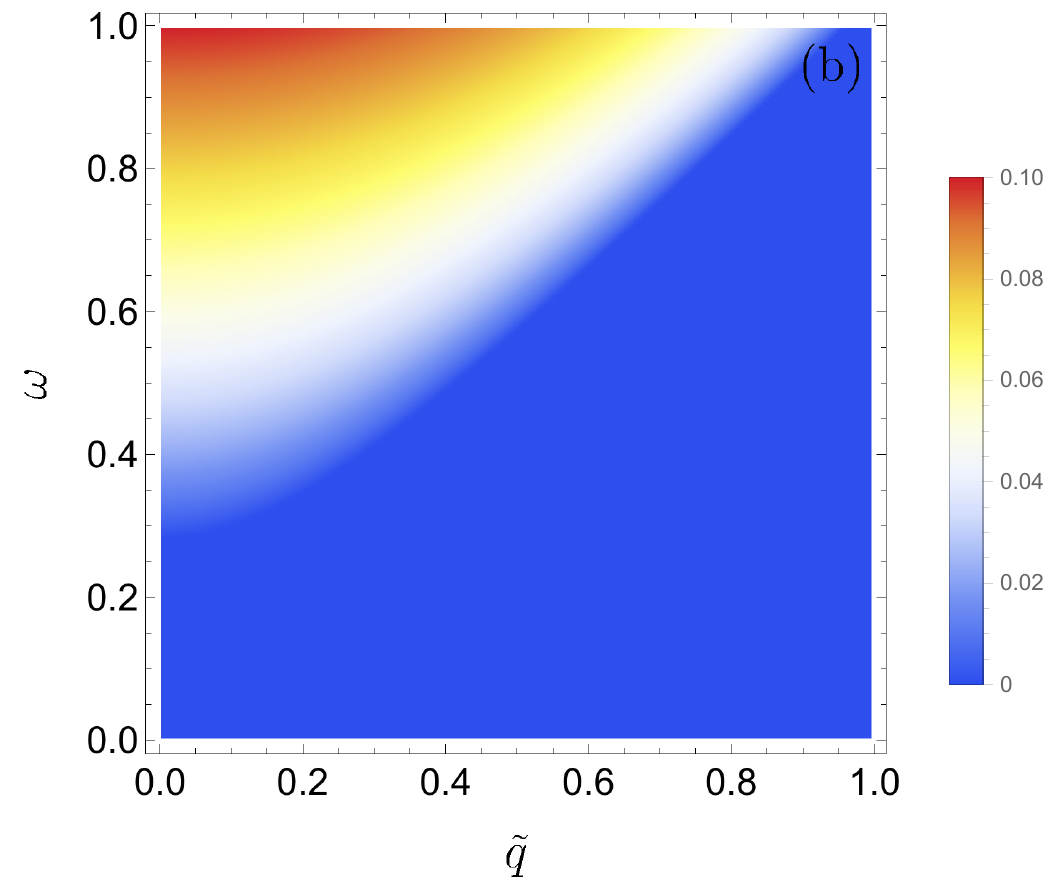}
		\caption{Imaginary parts of the polarization functions for $\lambda=0.1$, $m_x=0.1$, $m_y=0$ in the $\tilde{q}$-$\omega$ plane. (a) $-\im\Pi^{00}(\b{q},\omega)$, showing the gap $\sqrt{\tilde{q}^2+4\tilde{m}^2}$ and the optical absorption edge demarcating the white region from the red. (b) $\im\Pi^{xx}(\b{q},\omega)$.}
\label{fig:pol}
\vspace{-0.4cm}
\end{figure}

Eq.~\eqref{eq:cos} yields the useful identities at $\theta=\theta_0$:
\begin{eqnarray}
2\tilde{k}\tilde{q}\sin\theta_0&=&\sqrt{(\tilde{q}^2-\omega^2)[(\omega-2E_\b{k})^2-\tilde{q}^2]-4\tilde{q}^2\tilde{m}^2},\nonumber\\
\end{eqnarray}
and
\begin{eqnarray}
\tilde{\b{k}}\cdot(\tilde{\b{k}}+\tilde{\b{q}})+\tilde{m}^2&=&\frac{1}{2}\bigg((\omega-2E_\b{k})^2-\tilde{q}^2\bigg)+E_\b{k}(\omega-E_\b{k}),\nonumber\\
\end{eqnarray}
where we have switched to the dimensionful quantities $\tilde{k}\equiv v_Fk$, $\tilde{q}\equiv v_Fq$. From these, we get the following polarization components:
\begin{eqnarray}
\im\Pi^{0i}(\b{q},\omega)&=&-\frac{m_i}{\pi v_F^2}I_1 \nonumber\\
\im\Pi^{xy}(\b{q},\omega)&=&-\frac{2m_xm_y}{\pi v_F^2}I_2 \nonumber \\
\im\Pi^{ss}(\b{q},\omega)&=&-\frac{1}{\pi v_F^2}I_3+\frac{{\rm sgn}(s)}{\pi v_F^2}(\frac{1}{2}I_4+I_3)-\frac{2\tilde{m}_s^2}{\pi v_F^2}I_2,\nonumber\\
\end{eqnarray}
where we must evaluate the integrals
\begin{eqnarray}
I_1&=&\int_{\tilde{m}}^\infty dE_\b{k}\frac{2E_\b{k}-\omega}{\sqrt{(\tilde{q}^2-\omega^2)[(\omega-2E_\b{k})^2-\tilde{q}^2]-4\tilde{q}^2\tilde{m}^2}}\bigg|_\mathbb{R}\nonumber\\ \label{eq:im_int1}
I_2&=&\int_{\tilde{m}}^\infty dE_\b{k}\frac{1}{\sqrt{(\tilde{q}^2-\omega^2)[(\omega-2E_\b{k})^2-\tilde{q}^2]-4\tilde{q}^2\tilde{m}^2}}\bigg|_\mathbb{R}\nonumber\\ \label{eq:im_int2}
I_3&=&\int_{\tilde{m}}^\infty dE_\b{k}\frac{E_\b{k}(\omega-E_\b{k})}{\sqrt{(\tilde{q}^2-\omega^2)[(\omega-2E_\b{k})^2-\tilde{q}^2]-4\tilde{q}^2\tilde{m}^2}}\bigg|_\mathbb{R}\nonumber\\ \label{eq:im_int3}
I_4&=&\int_{\tilde{m}}^\infty dE_\b{k}\frac{(\omega-2E_\b{k})^2-\tilde{q}^2}{\sqrt{(\tilde{q}^2-\omega^2)[(\omega-2E_\b{k})^2-\tilde{q}^2]-4\tilde{q}^2\tilde{m}^2}}\bigg|_\mathbb{R}.\nonumber\\ \label{eq:im_int4}
\end{eqnarray}
The requirement that the square root be real further restricts the bounds of integration. The case $\tilde{q}<\omega$ yields the contradiction $E_\b{k}\leq\frac{1}{2}(\omega-\tilde{q}\sqrt{1+4\tilde{m}^2/(q^2-\omega^2)})<0$. On the other hand, the case $\omega>\tilde{q}$ yields the restriction
\be
(\omega-2E_\b{k})^2\leq\tilde{q}^2+\frac{4\tilde{q}^2}{\tilde{q}^2-\omega^2}.
\ee
This in turn implies
\be
\frac{\omega-\tilde{q}\sqrt{1+\frac{4\tilde{m}^2}{\tilde{q}^2-\omega^2}}}{2}\leq E_\b{k}\leq\frac{\omega+\tilde{q}\sqrt{1+\frac{4\tilde{m}^2}{\tilde{q}^2-\omega^2}}}{2},\label{eq:Ek_bounds}
\ee
and also $\omega\geq\sqrt{\tilde{q}^2+4\tilde{m}^2}$. Since the integration region is symmetric about $2E_\b{k}-\omega=0$, we see that $I_1=0$. The other integrals evaluate to
\begin{eqnarray}
I_2&=&\frac{\pi}{\sqrt{\omega^2-\tilde{q}^2}}\Theta(w-\sqrt{\tilde{q}^2+4\tilde{m}^2}) \nonumber \\
I_3&=&\frac{\pi}{16}\frac{(\omega^2-\tilde{q}^2)(\tilde{q}^2-2\omega^2)-4\tilde{q}^2\tilde{m}^2}{(\omega^2-\tilde{q}^2)^{3/2}}\Theta(w-\sqrt{\tilde{q}^2+4\tilde{m}^2})\nonumber\\
I_4&=&\frac{\pi}{4}\frac{\tilde{q}^2(\tilde{q}^2-\omega^2)-4\tilde{q}^2\tilde{m}^2}{(\omega^2-\tilde{q}^2)^{3/2}}\Theta(w-\sqrt{\tilde{q}^2+4\tilde{m}^2}),
\end{eqnarray}
which gives
\begin{eqnarray}
\im\Pi^{0i}(\b{q},\omega)&=&0 \nonumber \\
\im\Pi^{xy}(\b{q},\omega)&=&-\frac{2 m_xm_y}{v_F^2}\im d(\b{q},\omega)\nonumber\\
\im\Pi^{00}(\b{q},\omega)&=&\frac{1}{8v_F^2}\frac{\tilde{q}^2(\tilde{q}^2-\omega^2)-4\tilde{q}^2\tilde{m}^2}{(\omega^2-\tilde{q}^2)^{3/2}}\Theta(w-\sqrt{\tilde{q}^2+4\tilde{m}^2}) \nonumber \\
\im\Pi^{ii}(\b{q},\omega)&=&\frac{1}{v_F^2}\bigg(\frac{1}{4}\im c(\b{q},\omega)-2\tilde{m}_i^2\im d(\b{q},\omega)\bigg),
\end{eqnarray}
for $i\in\{x,y,z\}$, where 
\begin{eqnarray}
\im d(\b{q},\omega)&\equiv&\frac{1}{\sqrt{\omega^2-\tilde{q}^2}}\Theta(w-\sqrt{\tilde{q}^2+4\tilde{m}^2})\nonumber \\
\im c(\b{q},\omega)&\equiv&\sqrt{\omega^2-\tilde{q}^2}\Theta(w-\sqrt{\tilde{q}^2+4\tilde{m}^2}).
\end{eqnarray}
The imaginary parts of the $\Pi^{00}$ and $\Pi^{xx}$ polarization functions are shown in Fig.~\ref{fig:pol}. The gap of $\sqrt{\tilde{q}^2+4\tilde{m}^2}$ means that there is no Landau damping of any of the collective modes at low energies. For the plasmons, strong damping only occurs near the gap edge at large momenta.
\\

To evaluate the real parts, we use the Kramers-Kronig relation
\be
\re\Pi^{ss'}(\b{q},\omega)=\frac{2}{\pi}\mathcal{P}\int^\infty_0\frac{\omega'\im\Pi^{ss'}(\b{q},\omega')}{\omega'^2-\omega^2}d\omega'.\label{eq:KK}
\ee
For each case, the principal part must be taken when the pole at $\omega$ lies in the integration region, i.e. when $\omega\ge\sqrt{\tilde{q}^2+4\tilde{m}^2}$. Combining with the imaginary part we get.
\begin{eqnarray}
\Pi^{00}(\b{q},\omega)&=&\frac{1}{8v_F^2}\frac{\tilde{q}^2(\tilde{q}^2-\omega^2)-4\tilde{q}^2\tilde{m}^2}{(\tilde{q}^2-\omega^2)^{3/2}}\nonumber\\
&\times&\bigg(\frac{2}{\pi}\arctan\sqrt{\frac{4\tilde{m}^2}{\tilde{q}^2-\omega^2}}-1\bigg)-\frac{\tilde{q}^2\tilde{m}}{2\pi v_F^2(\tilde{q}^2-\omega^2)}\nonumber\\
\label{eq:pi00}\\
\Pi^{xy}(\b{q},\omega)&=&-\frac{2}{v_F^2}m_xm_yd(\b{q},\omega)=\Pi^{yx}(\b{q},\omega)\label{eq:pixy}\\
\Pi^{ii}(\b{q},\omega)&=&\frac{1}{v_F^2}\bigg(\frac{1}{4} c(\b{q},\omega)-2\tilde{m}_i^2 d(\b{q},\omega)\bigg),\label{eq:piii}
\end{eqnarray}
where
\be
d(\b{q},\omega)\equiv\frac{-1}{\sqrt{\tilde{q}^2-w^2}}\bigg(1-\frac{2}{\pi}\arctan\sqrt{\frac{4\tilde{m}^2}{\tilde{q}^2-w^2}}\bigg).
\ee
The result for the density-density channel was found for QED$_3$ in Ref.~\cite{appelquist1986}.
The first term $\im c(\b{q},\omega)$ in $\Pi^{ii}$, diverges for $\omega\rightarrow\infty$, which means the Kramers-Kronig relation is unusable. Instead, we evaluate this contribution directly from Eq.~\eqref{eq:piint}:
\begin{eqnarray}
c(\b{q},\omega)&=&8v_F^2\int\frac{d^2k}{(2\pi)^2}\sum_\alpha\frac{\alpha\bigg(1+\frac{\tilde{\b{k}}\cdot(\tilde{\b{k}}+\tilde{\b{q}})+\tilde{m}^2}{E_\b{k}E_{\b{k}+\b{q}}}\bigg)}{\omega+i\eta-\alpha(E_\b{k}+E_{\b{k}+\b{q}})}.\nonumber\\
\end{eqnarray}
The important fluctuations come from the static long-wavelength limit, so we first expand about $\omega=0$:
\begin{eqnarray}
c(\b{q},\omega)&=&\re c(\b{q},\omega)\\
&\approx& c(\b{q},0)\nonumber\\
&&-8v_F^2\int\frac{d^2k}{(2\pi)^2}\sum_\alpha\frac{E_\b{k}^2+E_\b{k}E_{\b{k}+\b{q}}+\tilde{\b{q}}\cdot\tilde{\b{k}}}{E_\b{k}E_{\b{k}+\b{q}}(E_\b{k}+E_{\b{k}+\b{q}})^3}\omega^2,\nonumber\\
\end{eqnarray}
where the term linear in $\omega$ vanished upon doing the sum on $\alpha$.
Expanding to second order about $q=0$ as well gives
\begin{eqnarray}
c(\b{q},\omega)&\approx&-16v_F^2\int\frac{d^2\b{k}}{(2\pi)^2}\frac{1}{E_\b{k}}-4v_F^2\int\frac{d^2\b{k}}{(2\pi)^2}\frac{1}{E_\b{k}^3}\omega^2\nonumber\\
&&-4v_F^2\int\frac{d^2\b{k}}{(2\pi)^2}\frac{3k^2\cos^2\theta-2E_\b{k}^2}{E_\b{k}^5}\tilde{q}^2,
\end{eqnarray}
where the term linear in $\tilde{q}$ vanishes upon doing the integral over $\theta$. Writing the first term as $c(0,0)$, we have
\begin{eqnarray}
c(\b{q},\omega)&\approx& c(0,0)+\frac{1}{\pi}\int^\infty_{\tilde{m}}dE_\b{k}\bigg(\frac{1}{E_\b{k}^2}+3\frac{\tilde{m}^2}{E_\b{k}^4}\bigg)\tilde{q}^2\nonumber\\
&&-\frac{2v_F^2}{\pi}\int^\infty_{\tilde{m}}\frac{dE_\b{k}}{2E_\b{k}^2}\omega^2\\
&=&c(0,0)+\frac{2}{\pi\tilde{m}}(\tilde{q}^2-\omega^2).
\end{eqnarray}
Note that the first term is divergent in the continuum model, but we can regularize it by expressing it in terms the lattice functions using $\sum_t\int\frac{d^2\b{k}}{(2\pi)^2}\rightarrow\frac{1}{N}\sum_\b{k}$.
\begin{eqnarray}
c(0,0)&\rightarrow&-\frac{8v_F^2}{N}\sum_\b{k}\frac{1}{E_\b{k}}\label{eq:c00}\\
&=&-8v_F^2\frac{4}{U},\label{eq:c002}
\end{eqnarray}
where the last equality holds for $m>0$ according to the saddle-point condition.
Likewise, expanding $d(\b{q},w)$ about $\omega=0$ gives 
\begin{eqnarray}
d(\b{q},\omega)&\approx&\frac{1}{\pi\tilde{q}}\bigg[-\pi+2\arctan\bigg(\frac{2\tilde{m}}{\tilde{q}}\bigg)\nonumber\\
&&+\bigg(-\frac{\pi}{2}+\frac{2\tilde{m}\tilde{q}}{4\tilde{m}^2+\tilde{q}^2}+\arctan\bigg(\frac{2\tilde{m}}{\tilde{q}}\bigg)\bigg)\frac{\omega^2}{\tilde{q}^2}\bigg].\nonumber\\
\end{eqnarray}
To second order in $\tilde{q}$, this becomes
\be
d(\b{q},\omega)\approx\frac{1}{\pi\tilde{m}}\bigg[-1+\frac{1}{12\tilde{m}^2}(\tilde{q}^2-\omega^2)\bigg].
\ee
So to second order in $\tilde{q}$ and $\omega$, the polarization components read
\begin{eqnarray}
\Pi^{00}(\b{q},\omega)&\approx&-\frac{1}{\pi v_F^2}\frac{\tilde{q}^2}{6\tilde{m}^2} +\mathcal{O}(\omega^2\tilde{q}^2)\label{eq:pi00_smallqw}\\
\Pi^{xy}(\b{q},\omega)&\approx&-\frac{2m_xm_y}{\pi v_F^2\tilde{m}}\bigg[-1+\frac{1}{12\tilde{m}^2}(\tilde{q}^2-\omega^2)\bigg]\\
\Pi^{ii}(\b{q},\omega)&\approx&-\frac{8}{U}+\frac{1}{v_F^2}\bigg(\frac{1}{2\pi\tilde{m}}(\tilde{q}^2-\omega^2)\nonumber\\
&& - \frac{2\tilde{m}_i^2}{\pi\tilde{m}}\bigg[-1+\frac{1}{12\tilde{m}^2}(\tilde{q}^2-\omega^2)\bigg]\bigg).\label{eq:piii_smallqw}
\end{eqnarray}
Note that the contribution from the charge fluctuations vanishes in the long wavelength limit.


The singularity of the fluctuation propagator at $q=\omega=0$ corresponds to the proliferation of fluctuations at the critical point of the magnetic transition. The condition $L^{-1}(0,0)=0$ is equivalent to the saddle-point condition that defines the critical coupling in Eq. (\ref{eq:Uc}). 

To see this, it is convenient to switch field variables from $x$ and $y$-channel fluctuations to amplitude and phase fluctuations, setting $\phi^x=m\cos\theta$ and $\phi^y=m\sin\theta$ with $m=\sqrt{m_x^2+m_y^2}$. Using $\delta\phi^x=\cos\theta\delta m-m\sin\theta\delta\theta$, and  $\delta\phi^y=\sin\theta\delta m+m\cos\theta\delta\theta$, we can identify the components of the fluctuation propagator in polar coordinates
\begin{eqnarray}
L^{-1}_{mm}&=&\cos^2\theta L^{-1}_{xx}+2\cos\theta\sin\theta L^{-1}_{xy}+\sin^2\theta L^{-1}_{yy}\\
L^{-1}_{m\theta}&=&2m(\cos(2\theta)L^{-1}_{xy}-\cos\theta\sin\theta L^{-1}_{xx}+\sin\theta\cos\theta L^{-1}_{yy})\nonumber\\
\\
L^{-1}_{\theta\theta}&=&m^2(\sin^2\theta L^{-1}_{xx}-2\sin\theta\cos\theta L^{-1}_{xy}+\cos^2\theta L^{-1}_{yy}).\nonumber\\
\end{eqnarray}
Using Eqs.~\eqref{eq:pixy},~\eqref{eq:piii}, we get
\begin{eqnarray}
L^{-1}_{mm}(\b{q},\omega)&=&1+\frac{U}{8v_F^2}\bigg(\frac{1}{4}c(\b{q},\omega)-2m^2d(\b{q},\omega)\bigg)\\\
L^{-1}_{m\theta}(\b{q},\omega)&=&0\\
L^{-1}_{\theta\theta}(\b{q},\omega)&=&m^2\bigg(1+\frac{U}{8v_F^2}\frac{c(\b{q},\omega)}{4}\bigg).
\end{eqnarray}
Using Eq.~\eqref{eq:c00} we note that to linear order in $m$, the condition $L^{-1}_{mm}(0,0)=0$ implies that 
\be
1-U\frac{1}{4N}\sum_\b{k}\frac{1}{\epsilon_\b{k}}=0,
\ee
which recovers the transition line $U=U_c$ in Eq. (\ref{eq:Uc}). 

More generally, to second order in $q$, and $\omega$, the inverse propagators on the Mott side of the transition ($m>0$) are
\begin{eqnarray}
L^{-1}_{zz}(\b{q},\omega)&\approx&\frac{U}{8\pi v_F^2\tilde{m}}\bigg[\pi\tilde{m}\bigg(\frac{8 v_F^2}{U}+\frac{c(0,0)}{4}\bigg)+2\lambda^2\nonumber\\
&&+\frac{1}{2}\bigg(1-\frac{\lambda^2}{3\tilde{m}^2}\bigg)(\tilde{q}^2-\omega^2)\bigg]\label{eq:Lz}\\
L^{-1}_{mm}(\b{q},\omega)&\approx&\frac{U}{8\pi v_F^2\tilde{m}}\bigg[\pi\tilde{m}\bigg(\frac{8 v_F^2}{U}+\frac{c(0,0)}{4}\bigg)+2m^2\nonumber\\
&&+\frac{1}{2}\bigg(1-\frac{m^2}{3\tilde{m}^2}\bigg)(\tilde{q}^2-\omega^2)\bigg]\\
L^{-1}_{\theta\theta}(\b{q},\omega)&\approx&m^2\frac{U}{8\pi v_F^2\tilde{m}}\bigg[\pi\tilde{m}\bigg(\frac{8v_F^2}{U}+\frac{c(0,0)}{4}\bigg)\nonumber\\
&&+\frac{1}{2}(\tilde{q}^2-\omega^2)\bigg].\label{eq:Ltheta}
\end{eqnarray}
Note that one cannot take the graphene limit ($\lambda\rightarrow0$) in these expressions for the spin fluctuation propagators, because in this limit, there are additional contributions to the polarization functions stemming from the mean-field $\phi^z$, which have not been included in this derivation. 

The poles of the propagators yield the spin-wave dispersions
\begin{eqnarray}
\omega_z(q)&=&\sqrt{v_F^2q^2+6\tilde{m}^2\bigg(\frac{2\lambda^2+\pi\tilde{m}(\frac{8v_F^2}{U}+\frac{c(0,0)}{4})}{3m^2+2\lambda^2}\bigg)}\nonumber\\
\\
\omega_m(q)&=&\sqrt{v_F^2q^2+6\tilde{m}^2\bigg(\frac{2m^2+\pi\tilde{m}(\frac{8v_F^2}{U}+\frac{c(0,0)}{4})}{3\lambda^2+2m^2}\bigg)}\nonumber\\
\\
\omega_\theta(q)&=&\sqrt{v_F^2q^2+2\pi\tilde{m}\bigg(\frac{8v_F^2}{U}+\frac{c(0,0)}{4}\bigg)}.
\end{eqnarray}
The phase fluctuations must lead to a gapless Goldstone mode due to spontaneous symmetry breaking of the $U(1)$ in-plane spin symmetry. We see that this only occurs if $\frac{c(0,0)}{4}=-\frac{8v_F^2}{U}$, which is precisely the saddle-point condition without Gaussian fluctuations (see Eqs.~\eqref{eq:c00}, \eqref{eq:c002}). That means that within this formalism, one cannot self-consistently include Gaussian fluctuations to modify the saddle-point condition without violating the Goldstone theorem. This problem is well-known in the context of the BCS-BEC crossover~\cite{diener2008, haussmann2007}. Nonetheless, we see that the phase fluctuations do not affect the transition line since their contribution to the free energy is minimized at the original mean-field value.

\begin{figure}[h]
		\centering
		\includegraphics[width=0.98\columnwidth]{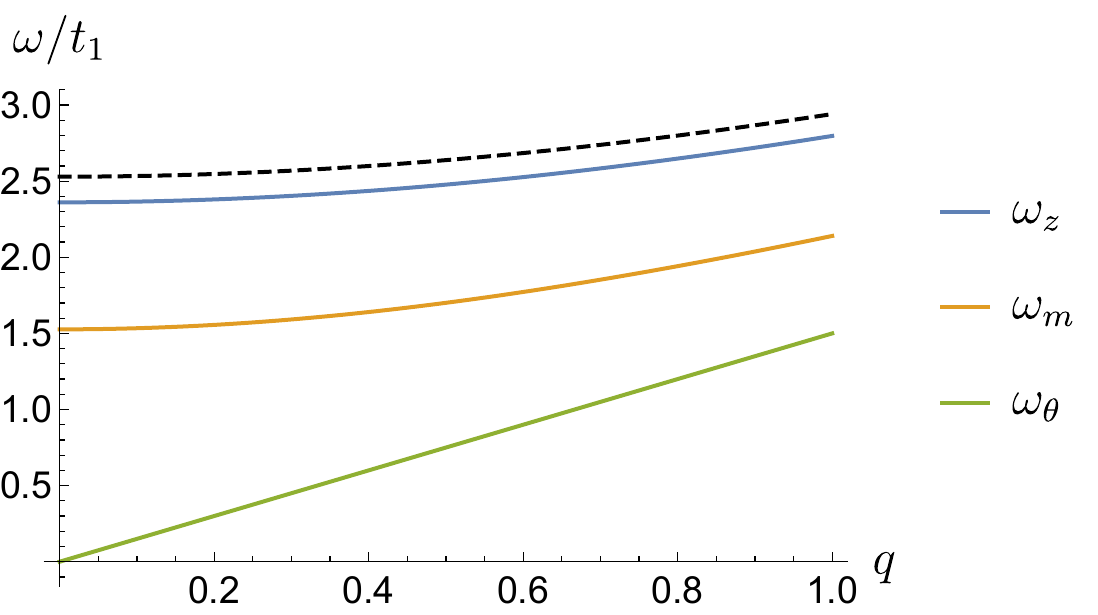}
		\caption{Collective spin-mode dispersions showing the out-of-plane and in-plane optical modes, and the acoustic Goldstone mode. The dashed black line shows the Landau damping edge above which the imaginary part of the polarization functions becomes non-zero. Here, the parameters are fixed deep in the Mott phase such that $t_2=0.2$ and $U=7$ for which the mean-field value of the in-plane magnetization is $\sqrt{(\phi^x)^2+(\phi^y)^2}\approx0.21$.}
\label{fig:dispersion}
\vspace{-0.4cm}
\end{figure}

If we stick with the original saddle-point condition, we see that there are two optical modes, one gapped by the spin-orbit coupling, and one gapped by the in-plane magnetic order parameter, as well as an acoustic Goldstone mode. Intriguingly, all of the excitations have the same velocity as the Dirac electrons. These dispersions, along with the Landau damping edge are shown in Fig.~\ref{fig:dispersion}.

The spin-wave dispersions will lead to zero-point fluctuations contributions as we discuss below showing the correspondence with the harmonic oscillator in three-dimensions. 
 
\subsection{Free energy}
The Gaussian spin fluctuations contribute an additional factor to the path integral
\be
\mathcal{Z}\sim m\bigg(\det(\beta UL^{-1}_{mm})\det(\beta UL^{-1}_{\theta\theta})\det(\beta UL^{-1}_{zz})\bigg)^{-1/2}.
\ee
This results in a shift in the free energy with respect to the mean-field value by an amount
\begin{eqnarray}
\Delta\mathcal{F}&=&\frac{1}{2\beta}\tr\ln(\beta UL^{-1}_{mm}\beta UL^{-1}_{\theta\theta}\beta UL^{-1}_{zz})\\
&\approx&\sum_{r=m,\theta,z}\frac{1}{2\beta}\int d^2\b{q}\sum_{i\omega_n}\ln(\alpha_r[(i\omega_n)^2-\omega_r^2(q)]),\nonumber\\
\end{eqnarray}
where we have Wick-rotated the retarded propagators approximated by the small $q$ and $\omega$ expressions in Eqs.~\eqref{eq:Lz}-\eqref{eq:Ltheta}. We have also defined
\begin{eqnarray}
\alpha_z&\equiv&-\frac{\beta U^2}{8\pi v_F^2\tilde{m}}\bigg(\frac{1}{2}-\frac{1}{3}\bigg(\frac{\lambda}{\tilde{m}}\bigg)^2\bigg)\\
\alpha_m&\equiv&-\frac{\beta U^2}{8\pi v_F^2\tilde{m}}\bigg(\frac{1}{2}-\frac{1}{3}\bigg(\frac{m}{\tilde{m}}\bigg)^2\bigg)\\
\alpha_\theta&\equiv&-\frac{\beta U^2}{16\pi v_F^2\tilde{m}}.
\end{eqnarray}

 The sum over $i\omega_n$ in the free energy runs over bosonic Matsubara frequencies. It is divergent, but this divergence is not physical and is easily regularized~\cite{leballac1996}. One way to do so is to write
\be
\frac{1}{\beta}\sum_{i\omega_n}\ln(\alpha_r[(i\omega_n)^2-\omega_r^2(q)]) =J^+_r+J^-_r + \frac{1}{\beta}\sum_{i\omega_n}\ln\alpha_r,
\ee
where 
\be
J_\pm^r\equiv\frac{1}{\beta}\sum_{i\omega_n}\ln(i\omega_n\pm\omega_r(q)).
\ee
The sum of $\ln\alpha_r$ is the divergent piece which can be removed. The remaining sums can be evaluated by noting that
\begin{eqnarray}
\frac{\partial J_\pm^r}{\partial\omega_r}&=&\frac{1}{\beta}\sum_{i\omega_n}\frac{\pm1}{i\omega_n\pm\omega_r(q)}\\
&=&\mp n_B(\mp\omega_r(q)),
\end{eqnarray}
where $n_B$ is the Bose distribution function. Integrating this gives
\begin{eqnarray}
J_+&=&\frac{1}{\beta}\ln(e^{\beta\omega_r(q)}-1)\\
J_-&=&-\omega_r(q)+\frac{1}{\beta}\ln(e^{\beta\omega_r(q)}-1),
\end{eqnarray}
where we have neglected constants independent of $\omega_r$. Thus,
\be
\Delta\mathcal{F}\approx\sum_{r=m,\theta,z}\int d^2\b{q}\frac{1}{2}\bigg(-\omega_r(q)+\frac{2}{\beta}\ln(e^{\beta\omega_r(q)}-1)\bigg).
\ee
In the zero-temperature limit, we recover
\be
\Delta\mathcal{F}\approx\sum_{r=m,\theta,z}\int d^2\b{q}\frac{1}{2}\omega_r(q),
\label{freeenergy}
\ee
which is just the zero-point energy of the Bose gas for each mode.

\section{Comparison of energetics}
\label{sec:HF}
An alternative way to compute the transition line is to decompose the interaction in Hartree-Fock-like terms:
\begin{eqnarray}
U \sum_{i} n_{i\uparrow} n_{i\downarrow} &\approx& U\sum_i\bigg[(\phi^x_i-i\phi^y_i)c^\dagger_{i\uparrow} c_{i\downarrow} +(\phi^x_i+i\phi^y_i)c^\dagger_{i\downarrow} c_{i\uparrow}\nonumber\\
&&-(\phi^0_i-\phi^z_i)n_{i\uparrow}-(\phi^0_i+\phi^z_i)n_{i\downarrow}-\b{\phi}_i\cdot\b{\phi}_i\bigg].\nonumber\\
\end{eqnarray}
Upon Fourier transforming this, and adding the non-interacting part $\mathcal{H}_{0,\b{k}}$, we obtain a new mean-field Hamiltonian $\mathcal{H}_{\mathrm{KMH}, \bm{k}}^{\mathrm{mf}}\equiv\mathcal{H}_{{0},\bm{k}}+\mathcal{H}_{U}(\b{\phi})$.
We introduce the unitary matrix $\mathcal{U}_\b{k}$ that diagonalizes $\mathcal{H}_{\mathrm{KMH}, \bm{k}}^{\mathrm{mf}}$ for a fixed set of parameters $\b{\phi}$ and fixed $\bm{k}$  as 
\begin{align}
		\Psi_{\bm{k}}^{\dagger} \mathcal{H}_{\mathrm{KMH}, \bm{k}}^{\mathrm{mf}} \Psi_{\bm{k}} &= \Psi_{\bm{k}}^{\dagger} \mathcal{U}_{\bm{k}}\mathcal{U}_{\bm{k}}^{\dagger} \mathcal{H}_{\mathrm{KMH},\bm{k}}^{\mathrm{mf}} \mathcal{U}_{\bm{k}}\mathcal{U}_{\bm{k}}^{\dagger} \Psi_{\bm{k}} \\
																							   &= \Omega_{\bm{k}}^{\dagger} \tilde{\mathcal{H}}_{\mathrm{KMH}, \bm{k}}^{\mathrm{mf}} \Omega_{\bm{k}}.
\end{align}
The new spinor basis is defined as $\Omega_{\bm{k}}\equiv\mathcal{U}_{\bm{k}}^{\dagger}\Psi_{\bm{k}}$, and the diagonal matrix as $\tilde{\mathcal{H}}_{\mathrm{KMH}, \bm{k}}^{\mathrm{mf}} \equiv \mathcal{U}_{\bm{k}}^{\dagger} \mathcal{H}_{\mathrm{KMH}, \bm{k}}^{\mathrm{mf}} \mathcal{U}_{\bm{k}}$.\\
Now we compute the amplitudes as

\begin{align}
		\left\langle c_{i\alpha}^{\dagger}c_{i\beta} \right\rangle 
																   &= \frac{1}{N} \sum_{\bm{k}} \sum_{\lambda,\lambda^{\prime}} \mathcal{U}_{\bm{k}\alpha\lambda}^{\ast} \mathcal{U}_{\bm{k}\beta\lambda^{\prime}}  \left\langle \Omega_{\bm{k}\lambda}^{\dagger} \Omega_{\bm{k}\lambda^{\prime}} \right\rangle \label{equ_diag_Omega} \\
																   &= \frac{1}{N} \sum_{\bm{k}}\sum_{\lambda\in{\rm occ}} \mathcal{U}_{\bm{k}\alpha\lambda}^{\ast} \mathcal{U}_{\bm{k}\beta\lambda}.
\end{align}
The composite indices $\alpha$ and $\beta$ run over sublattice and spin indices, 
while $\lambda$ in the last line only runs over occupied states.

We solve the self-consistent mean field equations 
for a given set of initial values for the fields $\b{\phi}$ by iteration. In each step, a new set of $\b{\phi}$ is computed from the previous set by computing the amplitudes 
$\left\langle c_{i\alpha}^{\dagger}c_{i\beta} \right\rangle$. This procedure is repeated until sufficient convergence is reached.\\

\begin{figure}[ht]
		\centering
		\includegraphics[width=0.98\columnwidth]{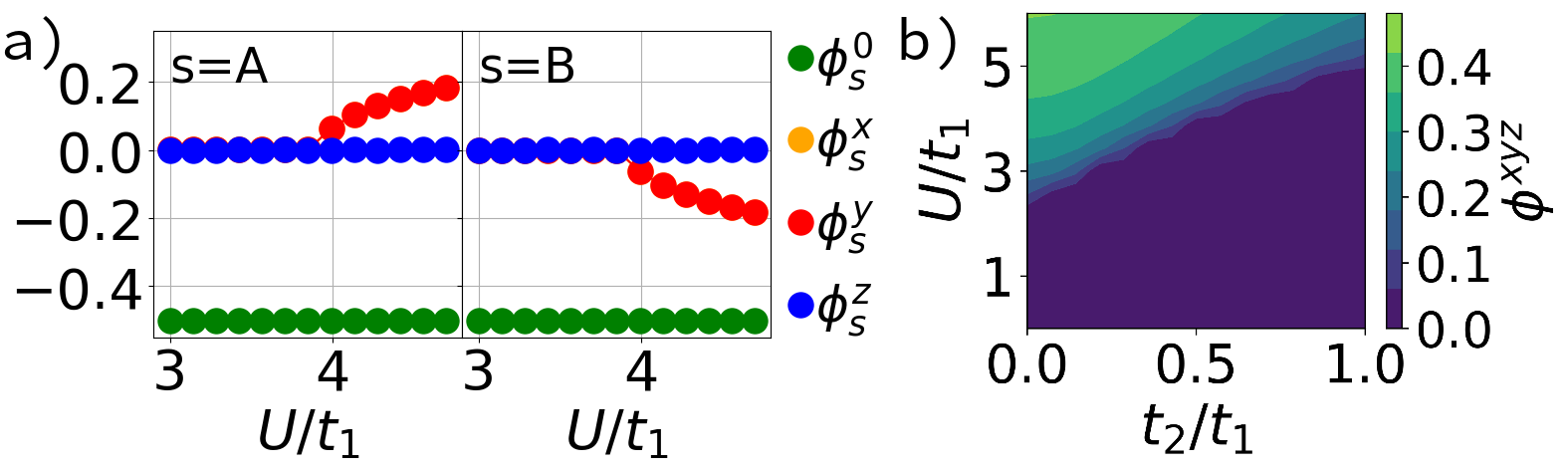}
		\caption{(a) Mean field values of the $\b{\phi}$ components on sublattices A (left) and B (right) for $t_2=0.5t_1$. $\phi^x$ and $\phi^y$ components are degenerate, here we just show one possible orientation of the in-plane field. (b) Magnitude of the combined magnetic order parameter in the $U$-$t_2$ plane.}
\label{fig_km_pd}
\vspace{-0.4cm}
\end{figure}

Fig.~\ref{fig_km_pd} shows the solution of the self-consistent mean field equations. Fig.~\ref{fig_km_pd} (a) verifies the emergence of the Mott phase with anti-ferromagnetic in-plane order. 
Fig.~\ref{fig_km_pd} (b) shows a two-dimensional $U-t_2$ phase diagram. Here, the quantity $\phi^{xyz} = \sqrt{\left(\phi^x\right)^2 + \left(\phi^y\right)^2 + \left(\phi^z\right)^2} $ captures the net magnetization. The position of the transition line differs with respect to the transition line obtained by the Green's function approach in Sec.~\ref{Results}. This is to be expected because although $\phi^0$, $\phi^x$, $\phi^y$, and $\phi^z$ are included in both methods, the choice of linear combination that makes up the mean fields is different as discussed in Sec.~\ref{app:methods}. 

\subsection{Note on the choice of mean field}\label{app:methods}

The transition line in Fig.~\ref{fig_km_pd}(b) differs from the transition line in Sec. \ref{Results} by a factor of two. This difference can be ascribed to a subtle difference in the choice of mean field. To see the difference we compute the mean-field free energy below in each case below. 

The choice in Sec.~\ref{sec:method} corresponds to a Heisenberg-like mean field that preserves the SU(2) symmetry of the interaction. We write the interaction as a ``spin'' Hamiltonian, where the spin vectors $\b{S}_i=c^\dagger_{i\alpha}\b{\sigma}_{\alpha\beta}c_{i\beta}$ form the natural mean fields:
\begin{eqnarray}
\mathcal{H}_U&=&\frac{U}{8}\sum_i\b{S}_i\cdot \b{S}_i +\frac{U}{4}\sum_i(n_{i\uparrow}+n_{i\downarrow})\\
&\approx&\frac{U}{8}\sum_i\bigg(\langle\b{S}_i\rangle\cdot\b{S}_i+\b{S}_i\cdot\langle\b{S}_i\rangle-\langle\b{S}_i\rangle\cdot\langle\b{S}_i\rangle+2\b{S}^0_i\bigg),\nonumber\\
\end{eqnarray}
where we have used the Minkowski inner product $\b{S}_i\cdot\b{S}_i=(S^0_i)^2-(S^x_i)^2-(S^y_i)^2-(S^z_i)^2$. Recalling that $\langle\b{S}_i\rangle=-2\b{\phi}_i$, we have
\be
\mathcal{H}_U\approx-\frac{U}{2}\sum_i\b{\phi}_i\cdot\b{S}_i-\frac{U}{2}\sum_i\b{\phi}_i\cdot\b{\phi}_i+\frac{U}{4}\sum_i \b{S}^0_i.
\ee
Fourier transforming gives
\be
\mathcal{H}_U\approx\sum_\b{k}\psi^\dagger_\b{k}\frac{U}{2}\mathcal{H}_{\rm int}\psi_\b{k}-\frac{UN}{2}(\b{\phi}_A\cdot\b{\phi}_A+\b{\phi}_B\cdot\b{\phi}_B),\label{eq:Sint}
\ee
where
\begin{eqnarray}
\mathcal{H}_{\rm int}&\equiv&\begin{pmatrix}
-\phi^0_A+\phi^z_A & 0 & \phi^x_A+i\phi^y_A & 0 \\
0 &-\phi^0_B+\phi^z_B & 0 & \phi^x_B+i\phi^y_B \\
\phi^x_A-i\phi^y_A & 0 & -\phi^0_A-\phi^z_A & 0 \\
0 & \phi^x_B-i\phi^y_B & 0 & -\phi^0_B-\phi^z_B
\end{pmatrix}\nonumber\\
&&+\frac{1}{2}\mathbb{I},
\end{eqnarray}
where $\mathbb{I}$ is the $4\times4$ identity matrix. For $\b{\phi}_A=-\b{\phi}_B$, we get the energy spectrum presented in Sec. \ref{Results}
\be
E_{\b{k}\pm}=\frac{U}{2}\pm\sqrt{\epsilon_\b{k}^2+2\gamma_\b{k}\bigg(\frac{U}{2}\bigg)\phi^z+\bigg(\frac{U}{2}\bigg)^2\vec\phi\cdot\vec\phi}.\label{eq:F_latt}
\ee
The free energy at half-filling ($\phi^0=-1/2$) is then
\begin{eqnarray}
F&=&UN-2\sum_\b{k}\sqrt{\epsilon_\b{k}^2+2\gamma_\b{k}\bigg(\frac{U}{2}\bigg)\phi^z+\bigg(\frac{U}{2}\bigg)^2\vec\phi\cdot\vec\phi}\nonumber\\
&& -UN\b{\phi}\cdot\b{\phi}\\
&=&\frac{3}{4}UN-2\sum_\b{k}\sqrt{\epsilon_\b{k}^2+2\gamma_\b{k}\bigg(\frac{U}{2}\bigg)\phi^z+\bigg(\frac{U}{2}\bigg)^2\vec\phi\cdot\vec\phi}\nonumber\\
&& + UN\vec{\phi}\cdot\vec{\phi}
\end{eqnarray}
so that
\be
\label{derivativef}
\frac{\partial F}{\partial\phi^x}=-2\sum_\b{k}\frac{(U/2)^2\phi^x}{\sqrt{\epsilon_\b{k}^2+2\gamma_\b{k}\bigg(\frac{U}{2}\bigg)\phi^z+\bigg(\frac{U}{2}\bigg)^2\vec\phi\cdot\vec\phi}}+2UN\phi^x.
\ee
Linearizing $\b{\phi}$ around the four-dimensional zero vector we get the minima condition
\be
\frac{1}{U_c}=\frac{1}{4N}\sum_\b{k}\frac{1}{\epsilon_\b{k}}.
\ee

On the other hand, the choice in Sec.~\ref{sec:HF} corresponds to a Hartree-Fock-like mean field decomposition where $\langle n_\sigma\rangle$, $\langle c^\dagger_\uparrow c_\downarrow\rangle$, $\langle c^\dagger_\downarrow c_\uparrow\rangle$ (which are linear combinations of the $\phi$'s)  form the natural mean fields:
\begin{eqnarray}
U \sum_{i} n_{i\uparrow} n_{i\downarrow} &\approx& U\sum_i\bigg[\langle n_\uparrow\rangle n_\downarrow+n_\uparrow\langle n_\downarrow\rangle -\langle c^\dagger_\uparrow c_\downarrow\rangle c^\dagger_\downarrow c_\uparrow\nonumber\\
&& -\langle c_\downarrow^\dagger c_\uparrow\rangle c_\uparrow^\dagger c_\downarrow -\langle n_\uparrow\rangle\langle n_\downarrow\rangle +\langle c_\uparrow^\dagger c_\downarrow\rangle\langle c_\downarrow^\dagger c_\uparrow\rangle\bigg]\nonumber\\
\\
&=& U\sum_i\bigg[(\phi^x_i+i\phi^y_i)c^\dagger_{i\uparrow} c_{i\downarrow} +(\phi^x_i-i\phi^y_i)c^\dagger_{i\downarrow} c_{i\uparrow}\nonumber\\
&&-(\phi^0_i-\phi^z_i)n_{i\uparrow}-(\phi^0_i+\phi^z_i)n_{i\downarrow}-\b{\phi}_i\cdot\b{\phi}_i\bigg].\nonumber\\
\end{eqnarray}
Fourier transforming gives
\begin{eqnarray}
U \sum_{i} n_{i\uparrow} n_{i\downarrow} &\approx&\sum_\b{k}\psi^\dagger_\b{k}U\left(\mathcal{H}_{\rm int}-\frac{1}{2}\mathbb{I}\right)\psi_\b{k}\nonumber\\
&&-UN(\b{\phi}_A\cdot\b{\phi}_A+\b{\phi}_B\cdot\b{\phi}_B).
\end{eqnarray}
Comparing to Eq.~\eqref{eq:Sint}, we see that this decomposition gives a mean-field interaction that is a factor of two larger than the Heisenberg-like decoupling.

For $\b{\phi}_A=-\b{\phi}_B$, we get the energy spectrum
\be
E_{\b{k}\pm}=-U\phi^0\pm\sqrt{\epsilon_\b{k}^2+2\gamma_\b{k}U\phi^z+U^2\vec\phi\cdot\vec\phi}.
\ee
The free energy at half-filling is then
\begin{eqnarray}
F&=&UN-2\sum_\b{k}\sqrt{\epsilon_\b{k}^2+2\gamma_\b{k}U\phi^z+U^2\vec\phi\cdot\vec\phi} -2UN\b{\phi}\cdot\b{\phi}\nonumber\\
\\
&=&\frac{1}{2}UN-2\sum_\b{k}\sqrt{\epsilon_\b{k}^2+2\gamma_\b{k}U\phi^z+U^2\vec\phi\cdot\vec\phi} +2UN\vec{\phi}\cdot\vec{\phi},\nonumber\\
\end{eqnarray}
so that
\be
\frac{\partial F}{\partial\phi^x}=-2\sum_\b{k}\frac{U^2\phi^x}{\sqrt{\epsilon_\b{k}^2+2\gamma_\b{k}U\phi^z+U^2\vec\phi\cdot\vec\phi}}+4UN\phi^x.
\ee
Linearizing $\b{\phi}$ about the zero vector we get the minima condition
\be
\frac{1}{U_c}=\frac{1}{2N}\sum_\b{k}\frac{1}{\epsilon_\b{k}}.
\ee

Both the Heisenberg-like and Hartree-Fock-like decompositions are perfectly valid choices of mean fields, but they lead to transition lines that differ by a factor of two. We have justified the choice in Eq.~\eqref{equ_decomposition_quartic_term} from the fact that it respects the spin-rotational symmetry of the Hubbard interaction and of the Heisenberg spin model in the Mott phase when $t_2\rightarrow 0$.

\section{Conclusion} 
\label{conclusion}
We develop a stochastic functional path integral approach from the variational principle to study the Mott transition in the interacting Kane-Mele model. We showed that by decomposing the interaction in an $SU(2)$ symmetric manner, an analytic transition line can be found that agrees quantitatively with numerical studies. The  magnetic ordering only occurs in-plane for any interaction strength, a result that was only previously established by spin models in the large $U$ limit. Our results indicate a second-order transition at the critical coupling. It should be noted that the Mott phase is described by a three-dimensional XY model~\cite{Lee_2011}, for which the mean-field magnetization profile is not expected to be accurate. However, in a multi-layer system, weak-coupling between the planes should be sufficient to stabilize the mean field behavior for the magnetism profile of Fig. \ref{fig:mag}. The method may be developed further to study fractional topological phases in systems with nearest-neighbour interactions, bilayers, as well as interacting topological superconductors. 

With regards to the topological number, we note that from the topological insulating phase, the ground state remains identical until the Mott transition as a result of $\phi^r=0$ (for all $r$). Therefore, the phase is entirely characterized by the $\mathbb{Z}_2$ topological invariant. Setting 
$\phi^r=0$, spin correlations then decay very rapidly, similarly as in a quantum spin liquid \cite{KMH1}. Using exact diagonalization in KWANT to study a ribbon geometry~\cite{groth2014}, we verify that the edge modes disappear in the Mott phase. This can be understood from the quantum field theory of the edge which is a Sine-Gordon model with gapped modes~\cite{KMH1}.

\acknowledgements
This work was supported jointly by the Natural Sciences and Engineering Research Council of Canada (NSERC) as well as the French ANR BOCA (JH and KLH) and the Deutsche Forschungsgemeinschaft (DFG, German Research Foundation) via Research Unit FOR 2414 under project number 277974659 (KLH). PWK also acknowledges Ecole Polytechnique for the support and funding for his PhD thesis. We also wish to thank Fakher Assaad for allowing us to compare with his QMC data.

\bibliographystyle{apsrev4-1}
\bibliography{bibi}
\end{document}